\documentclass[copyright,creativecommons]{eptcs}
\providecommand{\event}{SSV 2012} 
\usepackage{breakurl}             
\usepackage{wrapfig}
\usepackage{threeparttable}
\usepackage{amssymb,amsmath}
\usepackage{graphicx}
\usepackage[ruled,vlined]{algorithm2e}
\usepackage{url}
\usepackage{color}
\usepackage{txfonts}
\usepackage{multicol}

\providecommand{\urlalt}[2]{\href{#1}{#2}} 
\providecommand{\doi}[1]{doi:\urlalt{http://dx.doi.org/#1}{#1}}

\usepackage{wrapfig}
\usepackage{threeparttable}
\usepackage{amssymb,amsmath}
\usepackage{graphicx}
\usepackage[ruled,vlined]{algorithm2e}
\usepackage{url}
\usepackage{color}

\newsavebox{\tablebox}

\usepackage{xspace}

\newcommand{\true}{\texttt{True}}
\newcommand{\false}{\texttt{False}}

\newcommand\bool{\mathbb{B}}
\newcommand\boolform{{\cal B}}
\newcommand\eval{{\cal E}}
\newcommand{\initloc}[1]{l^0_{#1}}
\newcommand{\initeval}[1]{e^0_{#1}}

\newcommand{\system}{\mathcal{S}}
\newcommand{\Vis}{\textsf{Vis}}

\newcommand{\reach}{\mathcal{R}_{\system}\xspace}
\newcommand{\config}{\mathcal{C}_{\system}\xspace} 

\providecommand{\DontPrintSemicolon}{\dontprintsemicolon}

\newenvironment{list1}{\begin{list}{$\bullet$}
{\topsep 0 pt \parsep 0 pt \partopsep 0 pt \itemsep 0
pt}}{\end{list}}

\usepackage{xspace}
\newcommand{\Attr}{\mathsf{Attr}\xspace}

\newcommand{\sys}{\mathcal{S}}

\newtheorem{defi}{Definition}
\newtheorem{theo}{Theorem}
\newtheorem{prop}{Proposition}
\newtheorem{assump}{Assumption}

\SetAlgoSkip{} 

 \newcommand{\comment}[1]{}

\title{Distributed Priority Synthesis}
\author{Chih-Hong Cheng$^1$  \quad Rongjie Yan$^2$\quad Saddek Bensalem$^3$ \quad Harald Ruess$^1$
\institute{$^1$ Fortiss - An-Institut der TU M\"{u}nchen, Munich, Germany  \\
  $^2$ State Key Laboratory of Computer Science, Institute of Software, Beijing, China  \\
  $^3$ Verimag Laboratory, Grenoble, France}
\email{cheng@fortiss.org   \quad yrj@ios.ac.cn \quad saddek.bensalem@imag.fr \quad ruess@fortiss.org}
}
\def\titlerunning{Distributed Priority Synthesis}
\def\authorrunning{Cheng, Yan, Bensalem, Ruess}
\begin{document}
\maketitle

\begin{abstract}
Given a set of  interacting components with non-deterministic variable update and given  safety requirements, the goal  of {\em priority synthesis} is to restrict, by means of priorities,  the set of possible interactions in such a way as to guarantee the  given safety conditions for all possible runs. In {\em distributed priority synthesis} we are interested in obtaining local  sets of priorities, which
are deployed in terms of local component controllers sharing  intended next moves between  components in local neighborhoods only.
These possible communication paths between local controllers are specified  by means of a  {\em communication architecture}\@.  We formally define the problem of distributed priority synthesis in terms of a multi-player safety game between  players for (angelically) selecting the next transition of the components and an environment for (demonically) updating uncontrollable variables. We analyze the complexity of the problem, and propose several optimizations including a  solution-space exploration  based on  a diagnosis method using   a nested extension
 of the usual attractor computation in games together with a reduction to corresponding  SAT problems. When diagnosis fails, the method proposes potential candidates to guide the exploration. These optimized algorithms for solving  distributed priority synthesis problems have been integrated into the VissBIP framework. An experimental validation of this implementation  is performed  using  a range  of  case studies including scheduling in multicore processors and modular robotics.
\end{abstract}

\section{Introduction}

Distributed computing assemblies are usually built from interacting components with each component realizing a specific, well defined capability or service. Such a constituent component can be understood as a platform-independent computational entity that is described by means of its interface, which is published and advertised in the intended hosting habitat.

In effect, computing assemblies constrain the behavior of their constituent components to realize goal-directed behavior, and such a goal-directed orchestration of interacting components may be regarded
as synthesizing winning strategies in a multi-player game, with each constituent component and the environment a player. The game is won by the component players if the intended goals are achieved, otherwise the environment wins.
The orchestration itself may be centralized in one or several specialized controller components or the control may be distributed among the constituent components. Unfortunately, distributed controller
synthesis is known to be undecidable~\cite{PnueliFOCS90} in theory even
for reachability or simple safety conditions~\cite{Janin07On}. A number of decidable
subproblems have been proposed either by restricting the
communication structures between components, such as pipelined,
or by restricting the set of properties under consideration~\cite{madhusudan2002decidable,madhusudan2001distributed,mohalik:2003:distributed,finkbeiner2005uniform}.

In this paper we describe a solution to the distributed synthesis problem for automatically synthesizing local controllers which
are distributed among the constituent components. More precisely, given a set of interacting components with non-deterministic variable update and given a safety requirement on the overall system,
the goal of distributed priority synthesis is to restrict, by means of priorities on interactions, the set of possible interactions in such a way as to guarantee the given safety conditions. The structure of
these priorities is restricted in order to deploy the corresponding controllers in a distributed way, and communication between these local controllers is restricted based on a given communication architecture.

\begin{wrapfigure}{r}{0.5\textwidth}
    \centering
     \includegraphics[width=0.5\columnwidth]{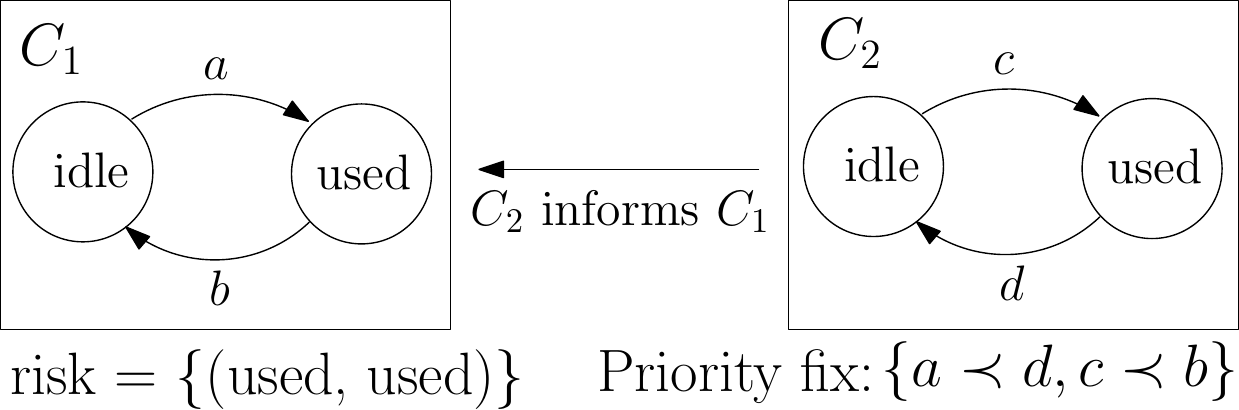}
      \caption{A sample example.}
     \label{fig:VissBIP.Sample}
\end{wrapfigure}
For example, Figure~\ref{fig:VissBIP.Sample} depicts two interacting components $C_1$
and $C_2$ with states \emph{idle} and \emph{used} and transitions $a$ through $d$ with
no further synchronization between the components. The goal is to
never simultaneously be in the risk state \emph{used}. This goal is achieved
by placing certain priorities on possible interactions. The priority
$a \prec d$, for example, inhibits transitions of $C_1$ from state \emph{idle} to
\emph{used}, whenever $C_2$ is ready to leave the state \emph{used}. This constraint
might be used as the basis of a local controller for $C_1$ as it is informed
by $C_2$ about its intended move using the given communication
channel.
Since many well-known scheduling strategies can be encoded by
means of priorities on interactions~\cite{goessler2003priority}, priority synthesis is closely
related to solving scheduling problems. In this way, the result of
distributed priority synthesis may also be viewed as a distributed
scheduler.

The rest of the paper is structured as follows. Section~\ref{sec.dps.formulation} contains background information on a simplified variant of the Behavior-Interaction-Priority (BIP) modeling framework~\cite{basu2006modeling}\@. The  corresponding  priority synthesis problem corresponds to  synthesizing a state-less winning strategy in a two-player safety game, where  the control player (angelically) selects the next transition of the components and the  environment player  (demonically) updates uncontrollable variables.
In Section \ref{sec.algo.prioritysyn.distributed.execution} we introduce the notion of deployable communication architectures and formally state the distributed priority synthesis problem. Whereas the general distributed controller synthesis problem is undecidable~\cite{PnueliFOCS90} we show that distributed priority synthesis is NP-complete\@. Overall, distributed priority synthesis is decidable over all communication architectures, as the methodology essentially searches for a strategy of a certain ``shape", where the shape is defined in terms of priorities. Section~\ref{sec.dps.algorithm} contains a solution to the distributed synthesis problem, which is guaranteed to be deployable on a given communication architecture. This algorithm is a generalization of the solution to the priority synthesis problem in~\cite{cheng:vissbip:2011,cheng:algo.priority.syn:2011}\@. 
It integrates essential optimizations based on symbolic game encodings including visibility constraints, followed by a nested attractor computation, and lastly, solving a corresponding (Boolean) satisfiability problem by extracting fix candidates while considering architectural constraints. Section~\ref{sec.dps.algo.extension}  describes some details and optimization of our implementation,
which is validated in Section~\ref{sec.dps.evaluation} against a set of selected case studies including scheduling in multicore processors and modular robotics. Section~\ref{sec.dps.related.work} contains related work and we conclude in Section~\ref{sec.dps.conclusion}.
Due to space limits, we leave proofs of propositions to our technical report~\cite{cheng:algo.dps:2011}.


\vspace{-3mm}
\section{Background}\label{sec.dps.formulation}
\vspace{-3mm}

Our notion of {\em interacting components} is heavily influenced by the Behavior-Interaction-Priority (BIP) framework~\cite{basu2006modeling} which consists of a set of automata (extended with data) that synchronize on joint labels; it is designed to model systems with combinations of synchronous and asynchronous composition.
For simplicity, we omit many syntactic features of BIP such as hierarchies of interactions and we restrict ourselves to Boolean data types only. Furthermore, uncontrollability is restricted to non-deterministic update of variables, and data transfer among joint  interaction among components is also omitted.

Let $\Sigma$ be  a nonempty alphabet of {\em interactions}\@.
A {\em component} $C_i$  of the form $(L_i, V_i$, $\Sigma_i, T_i, \initloc{i}$, $\initeval{i})$  is a {\em  transition system} extended with data, where  $L_i$  is a nonempty, finite set of \emph{control locations}, $\Sigma_i \subseteq \Sigma$ is  a nonempty subset of interaction labels used in $C_i$,  and $V_i$  is a finite set of \emph{(local) variables} of Boolean domain  $\bool = \{\true, \false\}$\@.
The set  $\eval(V_i)$ consists  of all evaluations $e: V_i\to\bool$\ over the variables $V_i$\@, and $\boolform(V_i)$ denotes the set of
 propositional formulas over variables in $V_i$; variable evaluations are extended to propositional formulas in the obvious way\@.
 $T_i$ is the set of \emph{transitions} of the form $(l,g,\sigma,f,l')$, where
                  $l, l'\in L_i$ respectively are the source and target locations,
                  the guard  $g \in \boolform(V_i)$  is a  Boolean formula over the variables $V_i$\@,
                  $\sigma \in \Sigma_i$ is an interaction label (specifying the event triggering the transition), and
                  $f: V_i \rightarrow (2^\bool \setminus \emptyset)$ 
                  is the {\em update relation} mapping every variable to a set of allowed Boolean values.
 Finally, $\initloc{i} \in L_i$ is the \emph{initial location} and $\initeval{i} \in \eval(V_i)$ is the initial evaluation of the variables.

\newcommand{\inter}{\bar{\sigma}}

A system $\mathcal{S}$ of {\em interacting components} is of the form  $(C, \Sigma, \mathcal{P})$, where $C=\{C_i\}_{1\leq i\leq m}$ is a set of  components,
the set of {\em priorities} $\mathcal{P} \subseteq 2^{\Sigma\times\Sigma}$ is irreflexive and transitive~\cite{goessler2003priority}.
 The notation  $\sigma_1 \prec \sigma_2$ is usually used instead of $(\sigma_1, \sigma_2) \in \mathcal{P}$, and we say that $\sigma_2$ has higher priority than $\sigma_1$\@.
A \emph{configuration (or state)} $c$ of a system $\mathcal{S}$ is of the form $(l_1, e_1, \ldots, l_m, e_m)$ with $l_i \in L_i$ and $e_i \in \eval(V_i)$ for all $i \in \{1,\ldots,m\}$\@.
The \emph{initial configuration} $c_0$ of   $\mathcal{S}$ is of the form  $(\initloc{1}, \initeval{1}, \ldots, \initloc{m}, \initeval{m})$\@.
An interaction $\sigma \in \Sigma$ is \emph{(globally) enabled} in a configuration $c$  if,
first, joint participation holds for $\sigma$, that is,
for all $i \in \{1,\ldots, m\}$, if $\sigma \in \Sigma_i$, then
there exists a transition  $(l_i,g_i,\sigma,f_i,l_i') \in T_i$ with  $e_i(g_i) = \true$\@,
and, second, there is no other interaction of higher priority for which joint participation holds.
$\Sigma_c$ denotes the set of (globally) enabled interactions in a configuration $c$\@.
For $\sigma \in  \Sigma_c$,
a configuration  $c'$ of the form $(l'_1, e'_1, \ldots, l'_m, e'_m)$ is a $\sigma$-\emph{successor} of $c$, denoted by  $c \xrightarrow[]{\sigma} c'$, if,
for all $i$ in $\{1,\ldots,m\}$:
 if $\sigma \not\in \Sigma_i$, then $l'_i = l_i$ and $e'_i = e_i$;
 if $\sigma \in \Sigma_i$  and (for some) transition
         of the form $(l_i,g_i,\sigma,f_i$, $l_i') \in T_i$ with $e_i(g_i) = \true$,
$e'_i= e_i[v_i / d_i]$ with $d_i \in f(v_i)$.

A  \emph{run} is of the form $c_0,\ldots,c_k$ with $c_0$ the initial configuration and   $c_j \xrightarrow[]{\sigma_{j+1}} c_{j+1}$  for all  $j: 0\le j< k$\@. In this case, $c_k$ is reachable, and  $\reach$  denotes the set of all reachable configurations from $c_0$\@.
Notice that such a sequence of configurations can be viewed  as an execution of a two-player game played alternatively between the control \textsf{Ctrl} and the environment \textsf{Env}\@.
In every position, player  \textsf{Ctrl} selects one of the enabled interactions and \textsf{Env} non-deterministically chooses new values for the variables before moving to the next position.
The game is won by \textsf{Env} if \textsf{Ctrl} is unable to select an enabled interaction, i.e., the system is deadlocked, or if  $\textsf{Env}$ is able to drive the run into a
bad configuration from some given set  $\mathcal{C}_{risk} \subseteq \config$\@.  More formally,  the system is \emph{deadlocked} in configuration
$c$ if there is no $c' \in \reach$ and no $\sigma \in \Sigma_c$ such that  $c \xrightarrow[]{\sigma} c'$, and the set of deadlocked states is denoted by $\mathcal{C}_{dead}$\@.
A configuration $c$ is \emph{safe} if $c \notin \mathcal{C}_{dead}\cup\mathcal{C}_{risk}$,
and a system is  safe if no reachable configuration is unsafe.
\begin{defi}[Priority Synthesis]
  Given a system $\mathcal{S} = (C , \Sigma, \mathcal{P})$ together with a set $\mathcal{C}_{risk} \subseteq \mathcal{C}_\sys$ of risk configurations,
 $\mathcal{P}_{+} \subseteq\Sigma\times\Sigma$ is a solution to the \emph{priority synthesis problem} if the extended
  system $(C , \Sigma, \mathcal{P}\cup\mathcal{P}_{+})$ is safe, and the defined relation of $\mathcal{P}\cup\mathcal{P}_{+}$ is also irreflexive and transitive.
\end{defi}

For the product graph induced by system $\system$, let $Q$ be the set of vertices and $\delta$ be the set of transitions.
In a single player game, where \textsf{Env} is restricted to deterministic updates, finding a solution to the priority synthesis problem is NP-complete in the size of $(|Q|+|\delta|+|\Sigma|)$~\cite{cheng:hardness:2011}\@.

\noindent \textbf{(Example in Fig.~\ref{fig:VissBIP.Sample})} The system $\mathcal{S}$ has two components $C_1, C_2$ (each component does not use any variable), uses interactions $\Sigma = \{a,b,c,d\}$, and has no predefined priorities. The initial configuration is $(idle, idle)$. Define the set of risk states to be $\{(used, used)\}$, then priority synthesis introduces $\{a\prec d, c\prec b\}$ as the set of priorities to avoid deadlock and risk states. Such a set ensures that whenever one component uses the resource, the other component shall wait until the resource is released. E.g., when $C_2$ is at $used$ and $C_1$ is at $idle$, priority $a \prec d$ can force $a$ to be disabled.

Examples of using non-controllable environment updates can be found in our extended report~\cite{cheng:algo.dps:2011}.

\section{Distributed Execution} \label{sec.algo.prioritysyn.distributed.execution}

We introduce the notion of (deployable) communication architecture for  defining distributed execution for a system  $\mathcal{S}$  of interacting components. Intuitively, a communication architecture specifies which components exchange information about their next intended move.

\newcommand{\informs}{\leadsto}

\begin{defi}
   A {\emph communication architecture} $Com$ for a system $\mathcal{S}$ of interacting components is  a set of ordered pairs of components of the form  $(C_i, C_j)$ for  $C_i, C_j \in C$\@.
   In this case we say that $C_i$  {\em informs} $C_j$ and we use the notation $C_i \informs C_j$\@.
   Such a  communication architecture $Com$ is  {\em deployable} if the following conditions hold for all $\sigma, \tau \in \Sigma$ and $i,j \in \{1,\ldots, m\}$\@:
  \begin{list1}
       \item {\em (Self-transmission)} $\forall i\in \{1,\ldots, m\}$, $C_i\informs C_i \in Com$.
    \item {\em (Group transmission)}  If $\sigma \in \Sigma_i \cap \Sigma_j$ then $C_j \informs C_i,~C_i \informs C_j \in Com$\@.
    \item {\em  (Existing priority transmission)} If $\sigma \prec \tau \in \mathcal{P}$,  $\sigma \in \Sigma_j$, and $\tau\in \Sigma_i$ then $C_i \informs C_j \in Com$\@.
  \end{list1}
\end{defi}
Therefore, components that possibly participate in a joint interaction exchange information about next intended moves (group transmission), and components with a high priority interaction $\tau$ need to inform all components with an interaction of lower priority than $\tau$ (existing priority transmission)\footnote{For the example in the introduction, to increase readability, we omit listing the communication structure for self-transmission and group transmission.}.
We make the following assumption.

\begin{assump}[Compatibility Assumption]
A system under synthesis has a deployable communication architecture.
\end{assump}

\noindent \textbf{(Example in Fig.~\ref{fig:VissBIP.Sample})} The communication architecture $Com$ in Fig.~\ref{fig:VissBIP.Sample} is $\{C_1 \informs C_1, C_2 \informs C_2, C_2 \informs C_1\}$. The original system $\mathcal{S}$ under $Com$ is deployable, but the modified system which includes the synthesized priorities $\{a\prec d, c\prec b\}$ is not, as it requires $C_1 \informs C_2$ to support the use of priority $c\prec b$ (when $C_2$ wants to execute $c$, it needs to know whether $C_1$ wants to execute $b$).

Next we define distributed notions of enabled interactions and behaviors, where all the necessary information is communicated along the defined communication architecture.

\begin{defi}
Given a communication architecture $Com$ for a system $\mathcal{S}$, an interaction $\sigma$ is \emph{visible} by $C_j$ if $C_i \informs C_j$
for all $i \in \{1,\ldots, m\}$ such that $\sigma\in\Sigma_i$\@.
Then for configuration $c=(l_1, e_1, \ldots, l_m, e_m)$, an interaction $\sigma \in \Sigma$ is \emph{distributively-enabled (at $c$)} if:
\begin{list1}
    \item (Joint participation: distributed version)
    for all $i$ with $\sigma \in \Sigma_i$: $\sigma$ is visible by $C_i$, and there exists  $(l_i,g_i,\sigma,\_,\_) \in T_i$ with  $e_i(g_i) = \true$\@.
    \item (No higher priorities enabled: distributed version)
    for all  $\tau \in \Sigma$ with  $\sigma \prec \tau$, and $\tau$ is visible by $C_i$:
    there is a $j\in \{1,\ldots, m\}$ such that $\tau \in \Sigma_j$ and either $(l_j,g_j,\tau,\_,\_) \not\in T_j$ or for every $(l_j,g_j,\tau,\_,\_) \in T_j$, $e_j(g_j) = \false$\@.
\end{list1}

\end{defi}

A configuration $c' = (l_1', e_1', \ldots,$ $l_m', e_m')$  is a {\em distributed $\sigma$-successor} of $c$ if $\sigma$ is distributively-enabled and $c'$ is a $\sigma$-successor of $c$.
 {\emph Distributed runs} are runs of system $\mathcal{S}$ under communication architecture $Com$\@.

Any move from a configuration to a successor configuration in the distributed semantics can be understood as a multi-player game with $(|C|+1)$ players between
controllers $\textsf{Ctrl}_i$ for each component and the external environment $\textsf{Env}$\@.
In contrast to the two-player game for the global semantics, $\textsf{Ctrl}_i$ now is only informed on the intended next moves of the components in the
visible region as defined by the communication architecture, and the control players play against the environment player.
First, based on the visibility, the control players agree (cmp. Assumption~\ref{assumption2} below) on an interaction $\sigma \in\Sigma_c$, and, second, the
environment chooses a $\sigma$-enabled transition for each component $C_i$ with $\sigma \in \Sigma_i$\@.
Now the successor state is obtained by local updates to the local configurations for each component and variables are non-deterministically
toggled by the environment\@.

\begin{prop}\label{prop.distributively.enableness}
Consider a system $\mathcal{S}=(C , \Sigma, \mathcal{P})$ under a deployable communication architecture $Com$. \emph{(a)} If $\sigma \in \Sigma$ is globally enabled at configuration $c$, then $\sigma$ is distributively-enabled at $c$. \emph{(b)} The set of distributively-enabled interactions at configuration $c$ equals $\Sigma_c$. \emph{(c)} If configuration $c$ has no distributively-enabled interaction, it has no globally enabled interaction.
\end{prop}

From the above proposition (part c) we can conclude that if configuration $c$ has no distributively-enabled interaction, then $c$ is deadlocked ($c\in \mathcal{C}_{dead}$)\@.
However we are looking for an explicit guarantee for the claim that the system at configuration $c$ is never deadlocked whenever there exists one distributively-enabled interaction in $c$\@. This means that whenever a race condition over a shared resource happens, it will be resolved (e.g., via the resource itself) rather than halting permanently and disabling the progress. Such an assumption can be fulfilled by variants of  distributed consensus algorithms such as majority voting (MJRTY)~\cite{boyer1991mjrty}.

\begin{assump}[Runtime Assumption] \label{assumption2}
 For a configuration $c$ with $|\Sigma_c| >0$, the distributed controllers $\textsf{Ctrl}_i$ agree on a distributively-enabled interaction $\sigma \in \Sigma_c$ for execution.
\end{assump}
The assumption assumes that the distributed semantics of a system can be implemented as the global semantics~\cite{Bonakdarpour2011distribute}.
With the above assumption, we then define, given a system $\mathcal{S}=(C , \Sigma, \mathcal{P})$ under a communication architecture $Com$, the set of deadlock states of $\mathcal{S}$ in distributed execution to be $\mathcal{C}_{dist.dead}=\{c\}$
where $c$ has no distributively-enabled interaction.
We immediately derive $\mathcal{C}_{dist.dead} = \mathcal{C}_{dead}$, as the left inclusion ($\mathcal{C}_{dist.dead} \subseteq \mathcal{C}_{dead}$) is the consequence of Proposition~\ref{prop.distributively.enableness}, and the right inclusion is trivially true.  With such an equality, given a risk configuration  $\mathcal{C}_{risk}$ and global deadlock states $\mathcal{C}_{dead}$,  we say that system $\mathit{S}$ under the distributed semantics
is \emph{distributively-safe} if there is no distributed run $c_0,\ldots,c_k$ such that $c_k\in \mathcal{C}_{dead}\cup\mathcal{C}_{risk}$;
a system that is not safe is called \emph {distributively-unsafe}\@.

\begin{defi}\label{defn.dps} 
  Given a system $\mathcal{S} = (C , \Sigma, \mathcal{P})$  together with a deployable communication architecture $Com$, the
  set of risk configurations $\mathcal{C}_{risk} \subseteq \mathcal{C}_\sys$,
  a set of priorities $\mathcal{P}_{d+}$  is a solution to the  {\em distributed priority synthesis problem}  if the following holds:
  1) $\mathcal{P}\cup\mathcal{P}_{d+}$ is transitive and irreflexive.
  2) $(C , \Sigma, \mathcal{P}\cup\mathcal{P}_{d+})$ is distributively-safe.
  3) For all  $ i,j \in \{1,\ldots, m\}$ s.t. $\sigma \in \Sigma_i$,  $\tau \in \Sigma_j$,
                  if $\sigma\prec\tau\in \mathcal{P}\cup\mathcal{P}_{d+}$ then  $C_j \informs C_i \in Com$.
\end{defi}
The 3rd condition states that newly introduced priorities are indeed deployable.
Notice that for system $\mathcal{S}$ with a deployable communication architecture $Com$,  and any risk configurations $\mathcal{C}_{risk}$ and global deadlock states $\mathcal{C}_{dead}$,
a solution to the distributed priority synthesis problem is distributively-safe iff it is (globally) safe.
Moreover, for a fully connected communication architecture, the problem of distributed priority synthesis reduces to (global) priority synthesis.

\begin{figure}
\centering
 \includegraphics[width=0.6\columnwidth]{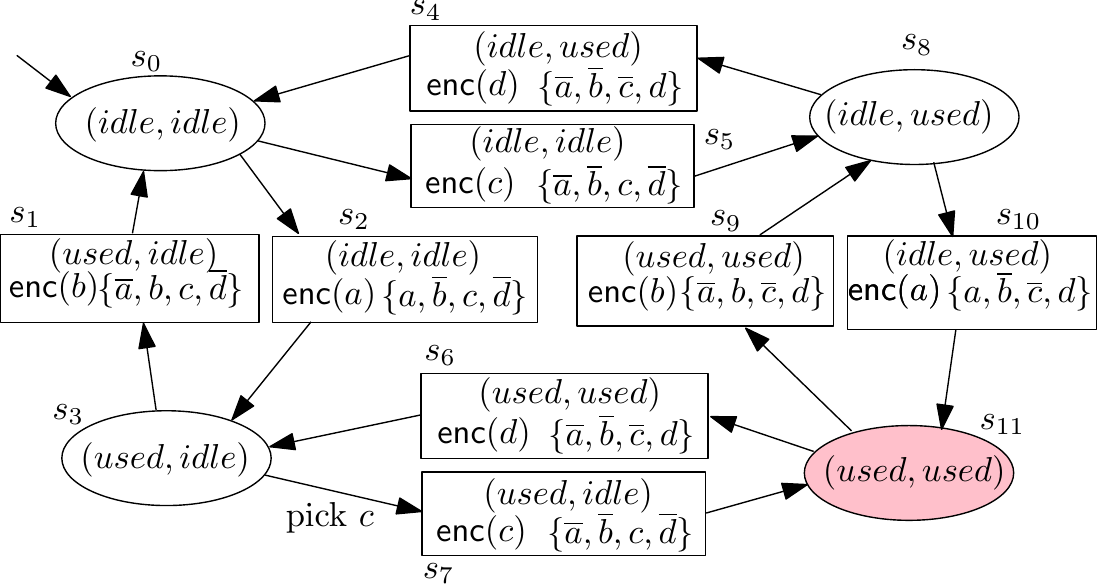}
  \caption{The symbolic encoding for the system in Fig.~\ref{fig:VissBIP.Sample}.}
 \label{fig:vissbip.encoding}
\end{figure}

\begin{theo}
Given system $\mathcal{S}=(C, \Sigma, \mathcal{P})$ under a deployable communication architecture $Com$, the problem of distributed priority synthesis is NP-complete to $|Q|+|\delta|+|\Sigma|$, where $|Q|$ and $|\delta|$ are the size of vertices and transitions in the product graph induced by $\system$\@, provided that $|C|^2 < |Q|+|\delta|+|\Sigma|$.
\end{theo}

\noindent (Sketch; see technical report~\cite{cheng:algo.dps:2011} for
full proof) First select a set of priorities (including $\mathcal{P}$)
and check if they satisfy transitivity, irreflexivity, architectural
constraints. Then check, in polynomial time, if the system under this
set of priorities can reach deadlock states; hardness follows from
hardness of global priority synthesis.

\noindent \textbf{(Example in Fig.~\ref{fig:VissBIP.Sample})} The
priority set $\{a \prec c, a\prec d\}$ is a feasible solution of
distributed priority synthesis, as these priorities can be supported
by the communication $C_2 \informs C_1$.

\newcommand{\enc}{enc}
\vspace{-4mm}
\section{Algorithmic Issues}\label{sec.dps.algorithm}
\vspace{-2mm}

It is not difficult to derive from the NP-completeness result
(Section~\ref{sec.algo.prioritysyn.distributed.execution}) a DPLL-like
search algorithm (see technical report~\cite{cheng:algo.dps:2011} for
such an algorithm), where each possible priority $\sigma \prec \tau$
is represented as a Boolean variable $\underline{\sigma \prec
  \tau}$. If $\underline{\sigma \prec \tau}$ is evaluated to \true,
then it is introduced in the priority. Then the algorithm checks if
such an introduced set is sufficient to avoid entering the
risk. Notice, however, that checking whether a risk state is reachable
is expensive\footnote{This suffers from the state-explosion
  problem. Therefore, if the size of the input is defined not to be
  the set of all reachable states but rather the number of components
  together with the size of each component, the problem is in
  PSPACE.}. As an optimization we therefore extend the basic search
algorithm above with a diagnosis-based fixing process.  In particular,
whenever the system is unsafe under the current set of priorities, the
algorithm diagnoses the reason for unsafety and introduces additional
priorities for preventing immediate entry into states leading to
unsafe states. If it is possible for the current scenario to be fixed,
the algorithm immediately stops and returns the fix. Otherwise, the
algorithm selects a set of priorities (from reasoning the inability of
fix) and uses them to guide the introduction of new priorities in the
search algorithm.


\begin{figure}
\begin{algorithm}[H]
\begin{scriptsize}
\DontPrintSemicolon
\SetKwInOut{Input}{input}\SetKwInOut{Output}{output}
\Input{System $\mathcal{S} = (C=(C_1,\ldots, C_m) , \Sigma, \mathcal{P})$, visibility constraint $\Vis_{\sigma_2}^{\sigma_1}$ where $\sigma_1,\sigma_2 \in \Sigma$}
\Output{Transition predicate $\mathcal{T}_{ctrl}$ for control and the set of deadlock states $\mathcal{C}_{dead}$}
\Begin{
     \texttt{let} predicate $\mathcal{T}_{ctrl} = \false$, $\mathcal{C}_{dead} := \true$\;
    \For{$\sigma\in \Sigma$}{
       \texttt{let} predicate $P_{\sigma} := \true$\;
     }
    \For{$\sigma\in \Sigma$}{
      \For{$i = \{1,\ldots,m\}$}{
        \nl\lIf{$\sigma \in \Sigma_i$}{
            $P_{\sigma} := P_{\sigma} \wedge \bigvee_{(l,g,\sigma,f,l')\in T_i} (\enc(l) \wedge g)$\;
         }
      }
    \nl $\mathcal{C}_{dead} := \mathcal{C}_{dead} \wedge \neg P_{\sigma}$\;
    }

    \For{$\sigma_1\in \Sigma$}{
    \nl \texttt{let} predicate $\mathcal{T}_{\sigma_1}:= p0 \wedge \neg p0' \wedge P_{\sigma_1} \wedge \textsf{enc}'(\sigma_1) \wedge \sigma_1'$\;
    \For{$\sigma_2 \in \Sigma, \sigma_2 \neq \sigma_1$}{
         \nl \lIf{$\Vis^{\sigma_2}_{\sigma_1} = \true$}{
          $\mathcal{T}_{\sigma_1} := \mathcal{T}_{\sigma_1} \wedge (P_{\sigma_2} \leftrightarrow \sigma_2')$\;
         }
         \nl\lElse{ $\mathcal{T}_{\sigma_1} := \mathcal{T}_{\sigma_1} \wedge \neg\sigma_2'$\;
         }
    }
    \For{$i = \{1,\ldots,m\}$}{
    \nl $\mathcal{T}_{\sigma_1}:= \mathcal{T}_{\sigma_1} \wedge \bigwedge_{y\in Y_i}  y \leftrightarrow y' \wedge \bigwedge_{v\in V_i} v \leftrightarrow v'$\;
    }
     \nl $\mathcal{T}_{ctrl}:= \mathcal{T}_{ctrl} \vee \mathcal{T}_{\sigma_1}$\;
    }
    \For{$\sigma_1 \prec \sigma_2 \in \mathcal{P}$}{
      \nl   $\mathcal{T}_{ctrl} := \mathcal{T}_{ctrl} \wedge ((\sigma_1' \wedge {\sigma_2'}) \to \neg {\textsf{enc}'(\sigma_1)});$ \;
     \nl         $\mathcal{T}_{12} =\mathcal{T}_{ctrl} \wedge (\sigma'_1 \wedge \sigma'_2);   \;
           \mathcal{T}_{ctrl} := \mathcal{T}_{ctrl} \setminus \mathcal{T}_{12}; $\;
   \nl         $ \mathcal{T}_{12,fix} := (\exists  \sigma'_1 : \mathcal{T}_{12}) \wedge (\neg \sigma'_1); $\;
    \nl        $ \mathcal{T}_{ctrl} := \mathcal{T}_{ctrl} \vee \mathcal{T}_{12, fix}$\;

    }
    \texttt{return} $\mathcal{T}_{ctrl}$, $\mathcal{C}_{dead}$\;
}
\caption{Generate controllable transitions and the set of deadlock states}\label{algo.prioritysyn.control.transition}
\end{scriptsize}
\end{algorithm}
\end{figure}

The diagnosis-based fixing process proceeds in two steps:

\noindent\textbf{(Step 1: Deriving fix candidates)} Game solving is used to derive potential fix candidates represented as a set of priorities.
    In the distributed case, we  need to encode visibility constraints: they specify for each interaction $\sigma$, the set of other interactions $\Sigma_{\sigma} \subseteq \Sigma$ visible to the components executing $\sigma$ (Section~\ref{subsec.dps.core.algorithm.encoding}). With visibility constraints, our game solving process results into a \emph{nested attractor computation} (Section~\ref{subsec.dps.core.algorithm.game}). %

\noindent\textbf{(Step 2: Fault-fixing)} Then create from fix candidates one feasible fix via solving a corresponding  SAT problem, which encodes properties of priorities and architectural restrictions (Section~\ref{subsec.dps.core.algorithm.SAT}).
If this propositional formula is unsatisfiable, then an  unsatisfiable  core is used to extract potentially useful  candidate  priorities.

\subsection{Game Construction}\label{subsec.dps.core.algorithm.encoding}

Symbolic encodings of interacting components form the basis of reachability checks, the diagnosis process, and  the algorithm for priority fixing (here we use $\mathcal{P}$ for $\mathcal{P}_{tran}$)\@.
In particular, symbolic encodings of system $\system = (C , \Sigma, \mathcal{P})$ use the following propositional variables:

\begin{list1}
    \item $p0$ indicates whether it is the controller's  or the  environment's turn.
    \item  $A=\{a_{1},\dots, a_{\lceil\log_{2}|\Sigma|\rceil}\}$ for the binary encoding $\textsf{enc}(\sigma)$ of the \emph{chosen interaction} $\sigma$  (which is agreed by distributed controllers for execution, see Assumption~\ref{assumption2}).
    \item $\bigcup_{\sigma \in \Sigma}\{\sigma\}$ are the
      variables representing interactions to encode \emph{visibility}.
      Notice that the same letter is used for an interaction and its corresponding encoding variable.
    \item $\bigcup_{i = 1}^m Y_i $, where
      $Y_i=\{y_{i1},\dots, y_{ik}\}$  for the binary encoding $\enc(l)$ of  locations $l \in L_i$\@.
   \item {\small $\bigcup_{i = 1}^m \bigcup_{v\in V_i} \{v\}$} are the encoding of the component variables.
\end{list1}
 Primed variables are used for encoding successor configurations and transition relations.
Visibility constraints  $\Vis^{\tau}_{\sigma} \in \{\true$, $\false\}$ denote the  visibility of interaction $\tau$ over another interaction $\sigma$\@.
It is computed statically: such a constraint  $\Vis^{\tau}_{\sigma}$ holds iff for $C_i, C_j \in C$ where $\tau \in \Sigma_i$ and $\sigma \in \Sigma_j $, $C_i\informs C_j \in Com$\@.

\noindent \textbf{(Example in Fig.~\ref{fig:VissBIP.Sample})} We
illustrate the symbolic game encoding using
Fig.~\ref{fig:vissbip.encoding} to offer an insight. Circles are
states which is the turn of the controller ($p0 = \true$) and squares
are those of the environment's turn ($p0 = \false$). The system starts
with state $s_0$ and proceeds by selecting an interaction that is
distributively enabled. E.g., $C_1$ may decide to execute $a$, and the
state then goes to $s_2$. In $s_2$, we have $\textsf{enc}(a)$ to
represent that $a$ is under execution. We also have $\{a,
\overline{b}, c, \overline{d}\}$ as the encoded visibility, as $C_2
\informs C_1$ and the availability of $c$ and $d$ can be passed. This
is used to represent that when $a$ is selected, $C_1$ is \emph{sure}
that $c$ is enabled at $C_2$. Then as no non-deterministic update is
in the system, the environment just moves to the successor by updating
the local state to the destination of the interaction $a$. Notice that
if $C_2$ decides to execute $c$, the play enters state $s_5$, which
has visibility $\{\overline{a}, \overline{b}, c, \overline{d}\}$. Such
a visibility reflects the fact that when $c$ executes, $C_2$ is
\emph{not aware of} the availability of $C_1$ to execute $a$, which is
due to the architectural constraint.


Following the above explanation,
Algorithms~\ref{algo.prioritysyn.control.transition}
and~\ref{algo.prioritysyn.environment.transition} return symbolic
transitions $\mathcal{T}_{ctrl}$ and $\mathcal{T}_{env}$ for the
control players $\bigcup_{i=1}^{m} \textsf{Ctrl}_i$ and the player
$\textsf{Env}$ respectively, together with the creation of a symbolic
representation $\mathcal{C}_{dead}$ for the deadlock states of the
system.  Line~1 of algorithm~\ref{algo.prioritysyn.control.transition}
computes when an interaction $\sigma$ is enabled.  Line~2 summarizes
the conditions for deadlock, where none of the interaction is
enabled. The computed deadlock condition can be reused throughout the
subsequent synthesis process, as introducing a set of priorities never
introduces new deadlocks.  In line~3, $\mathcal{T}_{\sigma_1}$
constructs the actual transition, where the conjunction with
$\textsf{enc}'(\sigma_1)$ indicates that $\sigma_1$ is the
\emph{chosen interaction} for execution.  $\mathcal{T}_{\sigma_1}$ is
also conjoined with $\sigma_1'$ as an indication that $\sigma_1$ is
enabled (and it can see itself).  Line~4 and~5 record the visibility
constraint. If interaction $\sigma_2$ is visible by $\sigma_1$
($\Vis^{\sigma_2}_{\sigma_1}=\true$), then by conjoining it with
$(P_{\sigma_2} \leftrightarrow \sigma_2')$, $\mathcal{T}_{\sigma_1}$
explicitly records the set of \emph{visible and enabled (but not
  chosen)} interactions. If interaction $\sigma_2$ is not visible by
$\sigma_1$, then the encoding conjuncts with $\neg \sigma_2'$.  In
this case $\sigma_2$ is \emph{treated as if it is not enabled} (recall
state $s_5$ in Fig.~\ref{fig:vissbip.encoding}): if $\sigma_1$ is a
bad interaction leading to the attractor of deadlock states, we cannot
select $\sigma_2$ as a potential escape (i.e., we cannot create
fix-candidate $\sigma_1 \prec \sigma_2$), as $\sigma_1 \prec \sigma_2$
is not supported by the visibility constraints derived by the
architecture.  Line~6 keeps all variables and locations to be the same
in the pre- and postcondition, as the actual update is done by the
environment.  For each priority $\sigma_1 \prec \sigma_2$, lines
from~8 to~11 perform transformations on the set of transitions where
both $\sigma_1$ and $\sigma_2$ are enabled.
Line~8 prunes out transitions from $\mathcal{T}_{ctrl}$ where both
$\sigma_1$ and $\sigma_2$ are enabled but $\sigma_1$ is chosen for
execution.  Then, the codes in lines~9 to~11 ensure that for remaining
transitions $\mathcal{T}_{12}$, they shall change the view as if
$\sigma_1$ is not enabled (line~10 performs the
fix). $\mathcal{T}_{ctrl}$ is updated by removing $\mathcal{T}_{12}$
and adding $\mathcal{T}_{12,fix}$.

\begin{figure}
\begin{algorithm}[H]
\begin{scriptsize}
\DontPrintSemicolon
\SetKwInOut{Input}{input}\SetKwInOut{Output}{output}
\Input{System $\mathcal{S} = (C=(C_1,\ldots, C_m) , \Sigma, \mathcal{P})$}
\Output{Transition predicate $\mathcal{T}_{env}$ for environment }
\Begin{
    \texttt{let} predicate $\mathcal{T}_{env} := \false$\;
    \For{$\sigma\in \Sigma$}{
        \texttt{let} predicate $T_{\sigma} := \neg p0 \wedge p0'$\;
         \For{$i = \{1,\ldots,m\}$}{
           \If{$\sigma \in \Sigma_i$}{
          \nl     $T_{\sigma} := T_{\sigma} \wedge \bigvee_{(l,g,\sigma,f,l')\in T_i}
    (\enc(l) \wedge g
    \wedge \enc'(l') \wedge  \textsf{enc}(\sigma) \wedge \textsf{enc}'(\sigma) \wedge \bigwedge_{v\in V_i} \cup_{e \in f(v)} v' \leftrightarrow e) $\;
            }
        }
        \For{$\sigma_1\in \Sigma, \sigma_1 \neq \sigma$}{
         \nl   $T_{\sigma} := T_{\sigma} \wedge \sigma_1' = \false$\;
        }
        \For{$i = \{1,\ldots,m\}$}{
           \nl \lIf{$\sigma \not\in \Sigma_i$}{
               $T_{\sigma} := T_{\sigma} \wedge \bigwedge_{y\in Y_i}  y \leftrightarrow y' \wedge \bigwedge_{v\in V_i} v \leftrightarrow v'$\;
            }
        }
        $\mathcal{T}_{env} := \mathcal{T}_{env} \vee T_{\sigma}$\;
    }
    \texttt{return} $\mathcal{T}_{env}$\;
}

\caption{Generate uncontrollable updates}\label{algo.prioritysyn.environment.transition}
\end{scriptsize}
\end{algorithm}
\end{figure}

\begin{prop} \label{prop.encoding.control}
Consider configuration $s$, where interaction $\sigma$ is (enabled and) chosen for execution. Given $\tau\in \Sigma$ at $s$ such that the encoding $\tau'=\true$ in Algorithm~\ref{algo.prioritysyn.control.transition}, then $\textsf{Vis}^{\tau}_{\sigma}=\true$ and interaction $\tau$ is also enabled at $s$.
\end{prop}
\begin{prop}\label{prop.encoding.deadlock}
 $\mathcal{C}_{dead}$  as returned by algorithm~\ref{algo.prioritysyn.control.transition} encodes the set of deadlock states of the input system $\mathcal{S}$\@.
\end{prop}

In Algorithm~\ref{algo.prioritysyn.environment.transition}, the environment updates the configuration using interaction $\sigma$ based on the indicator $\textsf{enc}(\sigma)$. Its freedom of choice in variable updates is listed in line~1 (i.e., $\cup_{e \in f(v)} v' \leftrightarrow e$). Line~2 sets all interactions $\sigma_1$ not chosen for execution to be false, and line~3 sets all components not participated in $\sigma$ to be stuttered.

\newcommand{\Occ}{\ensuremath{\textrm{Occ}}}
\newcommand{\Inf}{\ensuremath{\textrm{Inf}}}
\newcommand{\attr}{\ensuremath{\textsf{attr}}}
\newcommand{\N}{\mathbf{N}}

\begin{algorithm}[t]

\begin{scriptsize}
\DontPrintSemicolon
\SetKwInOut{Input}{input}\SetKwInOut{Output}{output}
\Input{Initial state $c_0$, risk states $\mathcal{C}_{risk}$, deadlock states $\mathcal{C}_{dead}$, set of reachable states $\reach(\{c_0\})$ and symbolic transitions $\mathcal{T}_{ctrl}$, $\mathcal{T}_{env}$ from Algorithm~\ref{algo.prioritysyn.control.transition} and~\ref{algo.prioritysyn.environment.transition} }
\Output{(1) Nested risk attractor $\textsf{NestAttr}_{env}(\mathcal{C}_{risk}\cup \mathcal{C}_{dead})$ and (2) $\mathcal{T}_{f} \subseteq \mathcal{T}_{ctrl}$, the set of control transitions starting outside $\textsf{NestAttr}_{env}(\mathcal{C}_{dead}\cup\mathcal{C}_{risk})$ but entering $\textsf{NestAttr}_{env}(\mathcal{C}_{risk}\cup \mathcal{C}_{dead})$.}
\Begin{
    \tcp{Create architectural non-visibility predicate}
    \nl\textbf{let} $\textsf{Esc} := \false$\;
    \For{$\sigma_i \in \Sigma$}{
        \nl\textbf{let} $\textsf{Esc}_{\sigma_i} := \textsf{enc}'(\sigma_i)$\;
         \nl \lFor{$\sigma_j \in \Sigma, \sigma_j \neq \sigma_i$}{
            $\textsf{Esc}_{\sigma_i} := \textsf{Esc}_{\sigma_i} \wedge \neg\sigma_j'$\;
         }
         $\textsf{Esc} := \textsf{Esc} \vee (\textsf{Esc}_{\sigma_i} \wedge \sigma_i')$\;
        }
    \tcp{Part A: Prune unreachable transitions and bad states, $\reach(\{c_0\})$ is the current set of reachable states}
    $\mathcal{T}_{ctrl} := \mathcal{T}_{ctrl} \wedge \reach(\{c_0\})$, $\mathcal{T}_{env} := \mathcal{T}_{ctrl} \wedge \reach(\{c_0\}); \;
    \mathcal{C}_{dead} := \mathcal{C}_{dead} \wedge \reach(\{c_0\})$, $\mathcal{C}_{risk} := \mathcal{C}_{risk} \wedge \reach(\{c_0\})$\;
    \tcp{Part B: Solve nested-safety game}
    \textbf{let} $\textsf{NestedAttr}_{pre} := \mathcal{C}_{dead} \vee \mathcal{C}_{risk}$,  $\textsf{NestedAttr}_{post} := \false$\;
    \nl\While{$\true$}{
    \tcp{B.1 Compute risk attractor of $\textsf{NestedAttr}_{pre}$}
    \nl\textbf{let} $\Attr := \textsf{compute\_risk\_attr}(\textsf{NestedAttr}_{pre}, \mathcal{T}_{env}, \mathcal{T}_{ctrl})$\;
    \tcp{B.2 Generate transitions with source in $\neg \Attr$ and destination in $\Attr$}
     \nl $\texttt{PointTo} := \mathcal{T}_{ctrl} \wedge \texttt{SUBS}((\exists \Xi': \Attr), \Xi, \Xi'))$\;
    \nl $\texttt{OutsideAttr} := \neg \Attr \wedge (\exists \Xi': \mathcal{T}_{ctrl})$\;
    \nl $\mathcal{T} := \texttt{PointTo} \wedge \texttt{OutsideAttr}$\;
    \tcp{B.3 Add the source vertex of B.2 to $\textsf{NestedAttr}$}
     \nl $\textsf{newBadStates} := \exists \Xi':(\mathcal{T} \wedge \textsf{Esc})$\;
     \nl $\textsf{NestedAttr}_{post} := \Attr \vee \textsf{newBadStates}$\;
    \tcp{B.4 Condition for breaking the loop}
     \lIf{$\textsf{NestedAttr}_{pre} \leftrightarrow \textsf{NestedAttr}_{post}$}{
        \texttt{break}\;
     }\lElse{
     $\textsf{NestedAttr}_{pre} := \textsf{NestedAttr}_{post}$\;
     }
    }
    \tcp{Part C: extract $\mathcal{T}_{f}$ }
    \nl $\texttt{PointToNested} := \mathcal{T}_{ctrl} \wedge \texttt{SUBS}((\exists \Xi': \textsf{NestedAttr}_{pre}), \Xi, \Xi'))$\;
    \nl $\texttt{OutsideNestedAttr} := \neg \textsf{NestedAttr}_{pre} \wedge (\exists \Xi': \mathcal{T}_{ctrl})$\;
    \nl $\mathcal{T}_{f} := \texttt{PointToNested} \wedge \texttt{OutsideNestedAttr}$\;
    \texttt{return} $\textsf{NestAttr}_{env}(\mathcal{C}_{dead}\cup\mathcal{C}_{risk}):=\textsf{NestedAttr}_{pre}$, $\mathcal{T}_{f}$\;
}

\caption{Nested-risk-attractor computation}\label{algo.prioritysyn.nested.attractor}
\end{scriptsize}
\end{algorithm}

\subsection{Fixing Algorithm:  Game Solving with Nested Attractor Computation}\label{subsec.dps.core.algorithm.game}
The first step of fixing is to compute the
\emph{nested-risk-attractor} from the set of bad states
$\mathcal{C}_{risk}\cup \mathcal{C}_{dead}$. Let $V_{ctrl}$
($\mathcal{T}_{ctrl}$) and $V_{env}$ ($\mathcal{T}_{env}$) be the set
of control and environment states (transitions) in the encoded
game. Let \emph{risk-attractor} $\Attr_{env}(X):=\bigcup_{k\in\N}
\attr^k_{env}(X)$, where $\attr_{env}(X) := X \cup \{ v\in V_{env}
\mid v\mathcal{T}_{env} \cap X \neq \emptyset \} \cup \{ v\in V_{ctrl}
\mid \emptyset \neq v\mathcal{T}_{ctrl} \subseteq X \}$, i.e.,
$\attr_{env}(X)$ extends state sets $X$ by all those states from which
either environment can move to $X$ within one step or control cannot
prevent to move within the next step. ($v\mathcal{T}_{env}$ denotes
the set of environment successors of $v$, and $v\mathcal{T}_{ctrl}$
denotes the set of control successors of $v$.)  Then
$\Attr_{env}(\mathcal{C}_{risk}\cup \mathcal{C}_{dead}) :=
\bigcup_{k\in\N} \attr^k_{env}(\mathcal{C}_{risk}\cup
\mathcal{C}_{dead})$ contains all nodes from which the environment can
force any play to visit the set $\mathcal{C}_{risk}\cup
\mathcal{C}_{dead}$.

\noindent \textbf{(Example in Fig.~\ref{fig:VissBIP.Sample})} Starting
from the risk state $s_{11}=(used, used)$, in attractor computation we
first add $\{s_{10}, s_{7}\}$ into the attractor, as they are
environment states and each of them has an edge to enter the
attractor. Then the attractor computation saturates, as for state
$s_3$ and $s_8$, each of them has one edge to escape from entering the
attractor. Thus $\Attr_{env}(\mathcal{C}_{risk}\cup
\mathcal{C}_{dead}) = \{s_{10}, s_{7}, s_{11}\}$.

Nevertheless, nodes outside the risk-attractor are not necessarily
safe due to visibility constraints.  We again use
Fig.~\ref{fig:vissbip.encoding} to illustrate such a concept. State
$s_3$ is a control location, and it is outside the attractor: although
it has an edge $s_3 \rightarrow s_7 $ which points to the
risk-attractor, it has another edge $s_3 \rightarrow s_1$, which does
not lead to the attractor. We call positions like $s_3$ as \emph{error
  points}. Admittedly, applying priority $c \prec b$ at $s_3$ is
sufficient to avoid entering the attractor. However, as $\Vis^{c}_{b}
= \false$, then for $C_2$ who tries to execute $c$, it is unaware of
the enableness of $b$. So $c$ can be executed freely by
$C_2$. Therefore, we should add $s_3$ explicitly to the (already
saturated) attractor, and \emph{recompute the attractor} due to the
inclusion of new vertices. This leads to an extended computation of
the risk-attractor (i.e., nested-risk-attractor).
\begin{defi}
The \emph{nested-risk-attractor} $\textsf{NestAttr}_{env}(\mathcal{C}_{risk}\cup \mathcal{C}_{dead})$  is the smallest superset of $\Attr_{env}(\mathcal{C}_{risk}\cup \mathcal{C}_{dead})$ such that the following holds.
\begin{enumerate}
\item For state $c \not\in \textsf{NestAttr}_{env}(\mathcal{C}_{risk}\cup \mathcal{C}_{dead})$, where there exists a (bad-entering) transition $t\in \mathcal{T}_{ctrl}$
with source $c$ and target $c' \in \textsf{NestAttr}_{env}(\mathcal{C}_{risk}\cup \mathcal{C}_{dead})$:
    \begin{list1}
    \item \emph{(Good control state shall have one escape)} there exists another transition $t'\in \mathcal{T}_{ctrl}$ such that its source is $c$ but its destination $c'' \not\in \textsf{NestAttr}_{env}(\mathcal{C}_{risk}\cup \mathcal{C}_{dead})$.
    \item\emph{ (Bad-entering transition shall have another visible candidate)} for every bad-entering transition $t$ of $c$, in the encoding let $\sigma$ be the chosen interaction for execution ($\textsf{enc}'(\sigma) = \true$). Then there exists another interaction $\tau$ such that, in the encoding, $\tau'=\true$.
    \end{list1}
\item \emph{(Add if environment can enter)} If $v\in V_{env}$, and $v\mathcal{T}_{env} \cap \textsf{NestAttr}_{env}(\mathcal{C}_{risk}\cup \mathcal{C}_{dead})  \neq \emptyset$, then $v \in \textsf{NestAttr}_{env}(\mathcal{C}_{risk}\cup \mathcal{C}_{dead})$.
\end{enumerate}
\end{defi}

Algorithm~\ref{algo.prioritysyn.nested.attractor} uses a nested
fixpoint for computing a symbolic representation of a nested risk
attractor.  The notation $\exists \Xi$ ($\exists \Xi'$) is used to
represent existential quantification over all umprimed (primed)
variables used in the system encoding.  Moreover, we use the operator
$\texttt{SUBS}(X, \Xi, \Xi')$, as available in many BDD packages, for
variable swap (substitution) from unprimed to primed variables in
$X$\@.  For preparation (line~1 to~3), we first create a predicate,
which explicitly records when an interaction $\sigma_i$ is enabled and
chosen (i.e., $\sigma_i' = \true$ and $\textsf{enc}'(\sigma_i) =
\true$). For every other interaction $\sigma_j$, the variable
$\sigma_j'$ is evaluated to $\false$ in BDD (i.e., either it is
disabled or not visible by $\sigma_i$, following
Algorithm~\ref{algo.prioritysyn.control.transition}, line~4 and~5).

The nested computation consists of two while loops (line~4,~5): B.1
computes the familiar risk attractor (see definition stated earlier,
and we refer readers to~\cite{cheng:algo.dps:2011} for concrete
algorithms), and B.2 computes the set of transitions $\mathcal{T}$
whose source is outside the attractor but the destination is inside
the attractor. Notice that for every source vertex $c$ of a transition
in $\mathcal{T}$: (1) It has chosen an interaction $\sigma \in \Sigma$
to execute, but it is a bad choice. (2) There exists another choice
$\tau$ whose destination is outside the attractor (otherwise, $c$
shall be in the attractor).  However, such $\tau$ may not be visible
by $\sigma$. Therefore, $\exists \Xi':(\mathcal{T} \wedge
\textsf{Esc})$ creates those states without any \emph{visible escape},
i.e., without any other visible and enabled interactions under the
local view of the chosen interaction. These states form the set of new
bad states $\textsf{newBadStates}$ due to architectural
limitations. Finally, Part~C of the algorithm extracts
$\mathcal{T}_{f}$ (similar to extracting $\mathcal{T}$ in B.2)\@.

Algorithm~\ref{algo.prioritysyn.nested.attractor} terminates, since
the number of states that can be added to $\textsf{Attr}$ (by
$\texttt{compute\_}$ $\texttt{risk\_attr}$) and
$\textsf{NestedAttr}_{post}$ (in the outer-loop) is finite.  The
following proposition is used to detect the infeasibility of
distributed priority synthesis problems.
\begin{prop}\label{prop.encoding.fast.diagnose}
Assume when executing the fix algorithm, only $\underline{\sigma\prec\tau}$, where $\sigma\prec\tau \in \mathcal{P}$, is evaluated to true.
If the encoding of the initial state is contained in $\textsf{NestAttr}_{env}(\mathcal{C}_{risk}\cup \mathcal{C}_{dead})$, then the
distributed priority synthesis problem for $\mathcal{S}$ with  $\mathcal{C}_{risk}$ is infeasible.
\end{prop}

\noindent \textbf{(Example in Fig.~\ref{fig:VissBIP.Sample})} We again
use the encoded game in Fig.~\ref{fig:vissbip.encoding} to illustrate
the underlying algorithm for nested-attractor computation. Line~5
computes the attractor $\{s_{10}, s_{7}, s_{11}\}$, and then Line~6 to
8 creates the set of transitions $\mathcal{T} = \{s_3 \rightarrow s_7,
s_8 \rightarrow s_{10}\}$, which are transitions whose source is
outside the attractor but the destination is inside. Remember that
$\textsf{Esc}$ in Line~2 generates a predicate which allows, when
$\sigma$ is enabled, only interaction $\sigma$ itself to be visible.
Then $\textsf{newBadStates} = \exists \Xi': \mathcal{T} \wedge
\textsf{Esc} = \exists \Xi': \{s_3 \rightarrow s_7\} = \{s_3\}$,
implying that such a state is also considered as bad. Then in Line~10
add $\{s_3\}$ to the new attractor and recompute. The computation
saturates with the following set of bad states $\{s_2, s_3, s_7,
s_{10}, s_{11}\}$. For state $s_0$ it has the risk edge $s_0
\rightarrow s_2$, but it can be blocked by priority $a\prec c$ (recall
in $s_2$ we have visibility $\{a,\overline{b},c,\overline{d}\}$). For
state $s_8$ it has the risk edge $s_8 \rightarrow s_{10}$, but it can
be blocked by priority $a\prec d$. Thus $\mathcal{T}_{f} = \{s_0
\rightarrow s_2, s_8 \rightarrow s_{10}\}$, and from $\mathcal{T}_{f}$
we can extract the candidate of the priority fix $\{a\prec c, a\prec
d\}$.

Notice that a visible escape is not necessarily a ``true escape" as
illustrated in Figure~\ref{fig:vissbip.locate.fix}. It is possible
that for state $c_2$, for $g$ its visible escape is $a$, while for $a$
its visible escape is $g$. Therefore, it only suggests candidates of
fixing, and in these cases, a feasible fix is derived in a SAT
resolution step (Section~\ref{subsec.dps.core.algorithm.SAT}).

\subsection{Fixing Algorithm: SAT Problem Extraction and Conflict Resolution}\label{subsec.dps.core.algorithm.SAT}

\begin{figure}
\centering
 \includegraphics[width=0.55\columnwidth]{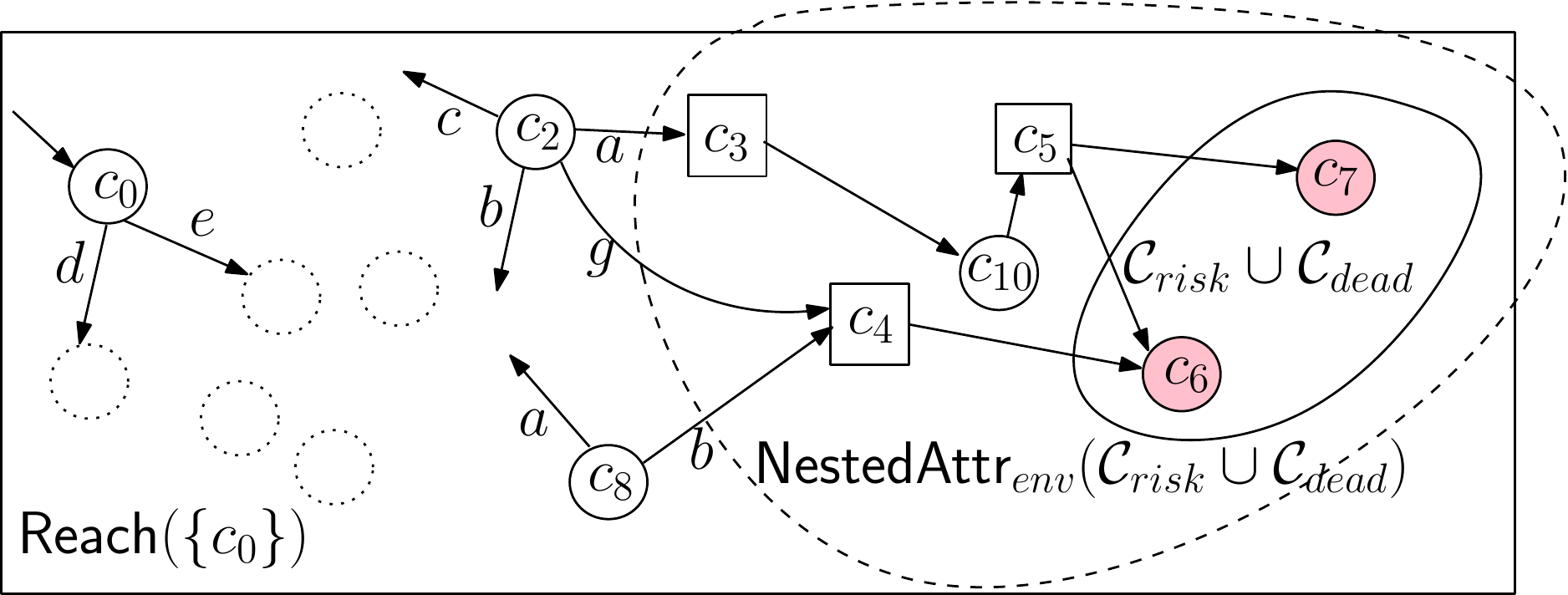}
  \caption{Locating fix candidates outside from the nested-risk-attractor.}
 \label{fig:vissbip.locate.fix}
\end{figure}

The returned value $\mathcal{T}_{f}$ of
Algorithm~\ref{algo.prioritysyn.nested.attractor} contains not only
the risk interactions but also all possible interactions which are
visible and enabled (see
Algorithm~\ref{algo.prioritysyn.control.transition} for encoding,
Proposition~\ref{prop.encoding.control} for result).  Consider, for
example, the situation depicted in Figure~\ref{fig:vissbip.locate.fix}
and assume that $\Vis^{c}_{a}$, $\Vis^{b}_{a}$, $\Vis^{c}_{b}$,
$\Vis^{a}_{g}$, and $\Vis^{a}_{b}$ are the only visibility constraints
which hold \true.  If $\mathcal{T}_{f}$ returns three transitions, one
may extract fix candidates from each of these transitions in the
following way.
\begin{list1}
    \item On $c_2$, $a$ enters the nested-risk-attractor, while $b,c$ are also visible from $a$; one obtains the  candidates $\{a \prec b, a \prec c\}$.
    \item On $c_2$, $g$ enters the nested-risk-attractor, while $a$ is also visible from $g$; one obtains the  candidate $\{g \prec a\}$.
    \item On $c_8$, $b$ enters the nested-risk-attractor, while $a$ is also visible; one obtains the candidate $\{b \prec a\}$.
\end{list1}
Using these candidates, one can perform \emph{conflict resolution} and
generate a set of new priorities for preventing entry into the
nested-risk-attractor region.  For example, $\{a\prec c, g \prec a, b
\prec a\}$ is such a set of priorities for ensuring the safety
condition.  Notice also that the set $\{a\prec b, g \prec b, b \prec
a\}$ is circular, and therefore not a valid set of priorities.

In our implementation, conflict resolution is performed using SAT
solvers.
Priority $\sigma_1 \prec \sigma_2$ is presented as a Boolean variable
$\underline{\sigma_1 \prec \sigma_2}$\@.  If the generated SAT problem
is satisfiable, for all variables $\underline{\sigma_1 \prec
  \sigma_2}$ which is evaluated to $\true$, we add priority $\sigma_1
\prec \sigma_2$ to the resulting introduced priority set
$\mathcal{P}_{d+}$.  The constraints below correspond to the ones for
global priority synthesis
framework~\cite{cheng:algo.priority.syn:2011}\@.
\begin{list1}
\item \emph{(1. Priority candidates)} For each edge $t \in
  \mathcal{T}_{f}$ which enters the risk attractor using $\sigma$ and
  having $ \sigma_1, \ldots, \sigma_e$ visible escapes (excluding
  $\sigma$), create clause $(\bigvee_{i=1}^e \underline{\sigma \prec
    \sigma_{i}})$.
\item \emph{(2. Existing priorities)} For each priority $\sigma \prec
  \tau \in \mathcal{P}$, create clause $(\underline{\sigma \prec
    \tau})$.
\item \emph{(3. Irreflexive)} For each interaction $\sigma$ used in
  (1) and (2), create clause $(\neg \underline{\sigma \prec \sigma})$.
\item \emph{(4. Transitivity)} For any $\sigma_1, \sigma_2, \sigma_3$
  used above, create a clause $((\underline{\sigma_1 \prec \sigma_2}
  \wedge \underline{\sigma_2 \prec \sigma_3})\Rightarrow
  \underline{\sigma_1 \prec \sigma_3})$.
\end{list1}
Clauses for architectural constraints also need to be added in the
case of distributed priority synthesis.  For example, if $\sigma_1
\prec \sigma_2$ and $\sigma_2 \prec \sigma_3$ then due to transitivity
we shall include priority $\sigma_1 \prec \sigma_3$. But if
$\Vis^{\sigma_3}_{\sigma_1} = \false$, then $\sigma_1 \prec \sigma_3$
is not supported by communication.  In the above example, as
$\Vis^{c}_{b} = \true$, $\{a\prec c, g \prec a, b \prec a\}$ is a
legal set of priority fix satisfying the architecture (because the
inferred priority $b \prec c$ is supported). Therefore, we introduce
the following constraints.
\begin{list1}
\item \emph{(5. Architectural Constraint)} Given $\sigma_1,\sigma_2\in \Sigma$, if $\Vis^{\sigma_2}_{\sigma_1} = \false$, then $\underline{\sigma_1 \prec \sigma_2}$ is evaluated to $\false$.
\item \emph{(6. Communication Constraint)} Given $\sigma_1,\sigma_2\in \Sigma$, if $\Vis^{\sigma_2}_{\sigma_1} = \false$, for any interaction $\sigma_3\in \Sigma$, if $\Vis^{\sigma_3}_{\sigma_1} = \Vis^{\sigma_2}_{\sigma_3} = \true$, at most one of $\underline{\sigma_1 \prec \sigma_3}$ or $\underline{\sigma_3 \prec \sigma_2}$ is evaluated to $\true$.
\end{list1}
A correctness argument of this fixing process  can be found   in our extended report~\cite{cheng:algo.dps:2011}.

\noindent \textbf{(Example in Fig.~\ref{fig:VissBIP.Sample})} By the nested attractor computation we have created a priority fix candidate $\{a\prec c, a\prec d\}$. Such a fix satisfies the above 6 types of constraints and thus is a feasible solution for distributed priority synthesis.

\section{Implementation}\label{sec.dps.algo.extension}
Our algorithm for solving the distributed priority synthesis problem
has been implemented in Java on top of the open-source workbench
VissBIP\footnote{\url{http://www6.in.tum.de/~chengch/vissbip}\@.}  for
graphically editing and visualizing systems of interacting
components. The synthesis engine itself is based on the JDD package
for binary decision diagrams, and the SAT4J propositional
satisfiability solver.  In addition, we implemented a number of
extensions and optimizations (e.g.,
Proposition~\ref{prop.encoding.fast.diagnose}) to the core algorithm
in Section~\ref{sec.dps.algorithm}; for lack of space details needed
to be omitted\@.

First, we also use the result of the unsatisfiable core during the fix
process to guide introducing new
priorities. 
E.g., if the fix does not succeed as both $\sigma \prec \tau$ and
$\tau \prec \sigma$ are used, the engine then introduces
$\underline{\sigma \prec \tau}$ for the next iteration. Then in the
next diagnosis process, the engine can not propose a fix of the form
$\tau \prec \sigma$ (as to give such a fix by the engine, it requires
that when $\tau$ and $\sigma$ are enabled while $\tau$ is chosen for
execution, $\sigma$ is also enabled; the enableness of $\sigma$
contradicts $\sigma \prec
\tau$). 

Second, we are over-approximating the nested risk attractor by
parsimoniously adding all source states in $\mathcal{T}_f$, as
returned from Algorithm~\ref{algo.prioritysyn.nested.attractor}, to
the nested-risk-attractor before recomputing; thereby increasing
chances of creating a new $\mathcal{T}_f$ where conflicts can be
resolved.

Lastly, whenever possible the implementation tries to synthesize local
controllers without any state information. If such a diagnosis-fixing
fails, the algorithm can also perform a model transformation of the
interacting components which is equivalent to transmitting state
information in the communication.
In order to minimize the amount of state information that is required
to communicate, we lazily extract refinement candidates from (minimal)
unsatisfiable cores of failed runs of the propositional solver, and
correspondingly refine the alphabet by including new state
information. Alternatively, a fully refined model transformation can
eagerly be computed in VissBIP\@.

\section{Evaluation}\label{sec.dps.evaluation}

We validate our algorithm using a collection of benchmarking models
including memory access problem, power allocation assurance, and
working protection in industrial automation; some of these case
studies are extracted from industrial case studies.
Table~\ref{tab:exp} summarizes the results obtained on an Intel
Machine with 3.4\,GHz Intel Core i7 CPU and 8\,GB RAM\@. Besides
runtime we also list the algorithmic extensions and optimizations
described in Section~\ref{sec.dps.algo.extension}\@.

The experiments 1.1 through 1.16 in Table~\ref{tab:exp} refer to
variations of the multiprocessor scheduling problem with increasing
number of processors and memory banks. Depending on the communication
architectures the engine uses refinement or extracts the UNSAT core to
find a solution.

Experiments 2.1 and 2.2 refer to a multi-robot scenario with possible
moves in a predefined arena, and the goal is to avoid collision by
staying within a predefined protection cap.  The communication
architecture is restricted in that the $i$-th robot can only notify
the $((i+1)\% n)$-th\@.  \comment{
\item \textbf{(Automation safety)} In an arena, every robot can make a
  set of possible movements to four directions. There exists more than
  one robot in the same arena. The risk in the current configuration
  is that all the robots may collapse in the same position. Our
  purpose is to form a rule-based protection cap to ensure that all
  the robots appearing simultaneously in the same location will never
  happen. Here robots are not fully informed, i.e., we explicitly
  define the communication structure as follows: $Robot[i]$ can notify
  $Robot[(i+1) \mod n]$.  }

In experiments 3.1 through 3.6 we investigate the classical dining
philosopher problem using various communication architectures.  If the
communication is clockwise, then the engine fails to synthesize
priorities\footnote{Precisely, in our model, we allow each philosopher
  to pass his intention over his left fork to the philosopher of his
  left. The engine uses Proposition~4 and diagnoses that it is
  impossible to synthesize priorities, as the initial state is within
  the nested-risk-attractor. }. If the communication is
counter-clockwise (i.e., a philosopher can notify its intention to his
right philosopher), then the engine is also able to synthesize
distributed priorities (for $n$ philosophers, $n$ rules
suffice). Compared to our previous priority synthesis technique, as in
distributed priority synthesis we need to separate visibility and
enabled interactions, the required time for synthesis is longer.

Experiment 4 is based on a case study for increasing the reliability
of data processing units (DPUs) by using multiple data sampling. The
mismatch between the calculated results from different devices may
yield deadlocks. The deadlocks can be avoided with the synthesized
priorities from VissBIP.

Finally, in experiment 5, we are synthesizing a decentralized
controller for the Dala robot~\cite{joser11}, which is composed of 20
different components. A hand-coded version of the control indeed did
not rule out deadlocks.  Without any further communication constraints
between the components, VissBIP locates the deadlocks and synthesizes
additional priorities to avoid them.

\begin{table*}[t]
\caption{Experimental results on distributed priority synthesis}\label{tab:exp}
\centering
\begin{scriptsize}
\begin{threeparttable}
\begin{tabular}{|l|l | c | c |c| l|}\hline
Index & Testcase and communication architecture & Components & Interactions & Time (seconds) & Remark\\\hline
1.1 & 4 CPUs with broadcast A & 8 &24 & 0.17 & x \\\hline
1.2 & 4 CPUs with local A, D & 8 &24 & 0.25 & A \\\hline
1.3 & 4 CPUs with local communication & 8 &24 & 1.66 & R \\\hline
1.4 & 6 CPUs with broadcast A & 12 &36 & 1.46 & RP-2 \\ \hline
1.5 & 6 CPUs with broadcast A, F & 12 &36 & 0.26 & x \\\hline
1.6 & 6 CPUs with broadcast A, D, F & 12 &36 & 1.50 & A \\\hline
1.7 & 6 CPUs with local communication  & 12 &36 & -  & fail \\\hline
1.8 & 8 CPUs with broadcast A & 16 &48 & 8.05 & RP-2 \\\hline
1.9 & 8 CPUs with broadcast A, H & 16 &48 & 1.30 & x \\\hline
1.10 & 8 CPUs with broadcast A, D, H & 16 & 48 & 1.80 & x \\\hline
1.11 & 8 CPUs with broadcast A, B, G, H & 16 & 48 & 3.88 & RP-2 \\\hline
1.12 & 8 CPUs with local communication & 16 & 48 & 42.80 & R \\\hline
1.13 & 10 CPUs with broadcast A & 20 & 60 & 135.03 & RP-2 \\\hline
1.14 & 10 CPUs with broadcast A, J & 20 & 60 & 47.89 & RP-2 \\\hline
1.15 & 10 CPUs with broadcast A, E, F, J  & 20 & 60 & 57.85 & RP-2 \\\hline
1.16 & 10 CPUs with local communication A, B, E, F, I, J  & 20 & 60 & 70.87 &RP-2 \\\hline\hline

2.1 & 4 Robots with 12 locations & 4 & 16 & 11.86 & RP-1 \\ \hline
2.2 & 6 Robots with 12 locations & 6 & 24 & 71.50 & RP-1 \\ \hline  \hline
3.1 & Dining Philosopher 10 (no communication) & 20 &  30 &  0.25 & imp \\ \hline
3.2 & Dining Philosopher 10 (clockwise next) & 20 &  30 &  0.27 & imp   \\ \hline
3.3 & Dining Philosopher 10 (counter-clockwise next) & 20 &  30 &  0.18 & x (nor: 0.16)  \\ \hline
3.4 & Dining Philosopher 20 (counter-clockwise next) & 40 &  60 &  0.85 & x,g (nor: 0.55)   \\ \hline
3.5 & Dining Philosopher 30 (counter-clockwise next) & 60 &  90 &  4.81 & x,g (nor: 2.75)  \\ \hline\hline
4 & DPU module (local communication) & 4 &  27 &   0.42 & x   \\ \hline\hline
5 & Antenna module (local communication) & 20 & 64 & 17.21 & RP-1 \\ \hline
\end{tabular}
\begin{tablenotes}
\item[x] Satisfiable by direct fixing (without assigning any priorities)
\item[A] Nested-risk-attractor over-approximation
\item[R] State-based priority refinement
\item[RP-1] Using UNSAT core: start with smallest amount of newly introduced priorities
\item[RP-2] Using UNSAT core: start with a subset of local non-conflicting priorities extracted from the UNSAT core
\item[fail] Fail to synthesize priorities (time out $> 150$ seconds using RP-1)
\item[imp] Impossible to synthesize priorities from diagnosis at base-level (using Proposition~4)
\item[g] Initial variable ordering provided (the ordering is based on breaking the circular order to linear order)
\item[nor] Priority synthesis without considering architectural constraints (engine in~\cite{cheng:algo.priority.syn:2011})
\end{tablenotes}
\end{threeparttable}
\end{scriptsize}
\end{table*}

\section{Related Work}\label{sec.dps.related.work}

Distributed controller synthesis is undecidable~\cite{PnueliFOCS90}
even for reachability or simple safety conditions~\cite{Janin07On}\@.
A number of decidable subproblems have been proposed either by
restricting the communication structures between components, such as
pipelined, or by restricting the set of properties under
consideration~\cite{madhusudan2002decidable,madhusudan2001distributed,mohalik:2003:distributed,finkbeiner2005uniform};
these restrictions usually limit applicability to a wide range of
problems\@.  Schewe and Finkbiner's~\cite{ScheweF07a} bounded
synthesis work on LTL specifications: when using automata-based
methods, it requires that each process shall obtain the same
information from the environment. The method is extended to encode
locality constraints to work on arbitrary structures. Distributed
priority synthesis, on one hand, its starting problem is a given
distributed system, together with an additional safety requirement
(together with the progress/deadlock-freedom property) to ensure. On
the other hand, it is also flexible enough to specify different
communication architectures between the controllers such as
master-slave in the multiprocessor scheduling example.  To perform
distributed execution, we have also explicitly indicated how such a
strategy can be executed on concrete platforms.

Starting with an arbitrary controller Katz, Peled and
Schewe~\cite{DBLP:conf/cav/KatzPS11,DBLP:conf/atva/KatzPS11} propose a
knowledge-based approach for obtaining a decentralized controller by
reducing the number of required communication between components. This
approach assumes a fully connected communication structure, and the
approach fails if the starting controller is inherently
non-deployable.

Bonakdarpour, Kulkarni and Lin~\cite{bonakdarpour2011automated}
propose methods for adding fault-recoveries for BIP components.  The
algorithms in~\cite{bonakdarpour2008sycraft,bonakdarpour2011automated}
are orthogonal in that they add additional behavior, for example new
transitions, for individual components instead of determinizing
possible interactions among components as in distributed priority
synthesis\@.  However, distributed synthesis described by Bonakdarpour
et al.~\cite{bonakdarpour2008sycraft} is restricted to local processes
without joint interactions between components\@.

Lately, the problem of deploying priorities on a given architecture
has gained increasing
recognition~\cite{Bonakdarpour2011distribute,bensalem2010methods}; the
advantage of priority synthesis is that the set of synthesized
priorities is always known to be deployable.

\section{Conclusion}\label{sec.dps.conclusion}

We have presented a solution to the distributed priority synthesis
problem for synthesizing deployable local controllers by extending
the 
algorithm for synthesizing stateless winning strategies in safety
games~\cite{cheng:vissbip:2011,cheng:algo.priority.syn:2011}\@.  We
investigated several algorithmic optimizations and validated the
algorithm on a wide range of synthesis problems from multiprocessor
scheduling to modular robotics.  Although these initial experimental
results are indeed encouraging, they also suggest a number of further
refinements and extensions.

The model of interacting components can be extended to include a rich
set of data types by either using Boolean abstraction in a
preprocessing phase or by using satisfiability modulo theory (SMT)
solvers instead of a propositional satisfiability engine; in this way,
one might also synthesize distributed controllers for real-time
systems.  Another extension is to to explicitly add the faulty or
adaptive behavior by means of demonic non-determinism.

Distributed priority synthesis might not always return the most useful
controller. For example, for the Dala robot, the synthesized
controllers effectively shut down the antenna to obtain a
deadlock-free system. Therefore, for many real-life applications we
are interested in obtaining optimal, for example wrt. energy
consumption, or Pareto-optimal controls.

Finally, the priority synthesis problem as presented here needs to be
extended to achieve goal-oriented orchestration of interacting
components. Given a set of goals in a rich temporal logic and a set of
interacting components, the orchestration problem is to synthesize a
controller such that the resulting assembly of interacting components
exhibits goal-directed behavior. One possible way forward is to
construct bounded reachability games from safety games.

Our vision for the future of programming is that, instead of
painstakingly engineering sequences of program instructions as in the
prevailing Turing tarpit, designers rigorously state their intentions
and goals, and the orchestration techniques based on distributed
priority synthesis construct corresponding goal-oriented assemblies of
interacting components~\cite{wegner1997interaction}.

\section{Acknowledgement}
Chih-Hong Cheng thanks Christian Buckl and Alois Knoll for their early
feedback on this work.

\bibliographystyle{eptcs}

\end{document}